\documentclass[final,5p,authoryear,times]{elsarticle}
\pdfoutput=1

\usepackage{graphicx}
\usepackage{amsmath}
\usepackage{booktabs}
\usepackage{colortbl} 
\usepackage{amssymb}
\usepackage{makecell}
\usepackage{color}

\usepackage[T1]{fontenc}
\usepackage{ae,aecompl}
\usepackage{enumitem}
\usepackage{times}
\usepackage[hidelinks]{hyperref}
\usepackage{xcolor}
\hypersetup{
    colorlinks,
    linkcolor={red!50!black},
    citecolor={blue!50!black},
    urlcolor={blue!80!black}
}
\usepackage{multirow}
\usepackage{mwe}
\usepackage{rotating}
\usepackage[normalem]{ulem}
\useunder{\uline}{\ul}{}
\usepackage{dblfloatfix}

\newcommand{\rthis}[1]{\textcolor{black}{#1}}
\newcommand{\bthis}[1]{\textcolor{black}{#1}}
\journal{Astronomy \& Computing}
\begin{document}\sloppy
\begin{frontmatter}

\title{Separation of pulsar signals from noise using supervised machine learning algorithms}

\author[1]{Suryarao Bethapudi}\ead{ep14btech11008@iith.ac.in}
\author[1]{Shantanu  Desai}\ead{shantanud@iith.ac.in}
\address[1]{Department of Physics, IIT Hyderabad, Kandi,  Telangana-502285, India}




\begin{abstract}
We evaluate  the performance of four different machine learning (ML) algorithms: an Artificial Neural Network Multi-Layer Perceptron (ANN MLP), Adaboost, Gradient Boosting Classifier (GBC), and XGBoost, for  the separation of pulsars
from radio frequency interference (RFI) and other sources of noise, using a dataset  obtained from the post-processing  of a pulsar search pipeline. This dataset was previously used for the cross-validation of the {\tt SPINN}-based  machine learning engine, obtained  from the  reprocessing
 of the     HTRU-S survey data~\citep{Morello}. We have used the Synthetic Minority Over-sampling Technique (SMOTE) to deal with high-class imbalance
 in the dataset. We report a variety of quality  scores from all four of these algorithms 
on both  the non-SMOTE and  SMOTE datasets. 
For all the above ML methods, we report high accuracy and G-mean for  both the non-SMOTE and SMOTE cases. 
We study the feature importances using Adaboost, GBC, and XGBoost and also from the minimum Redundancy Maximum Relevance  approach to report algorithm-agnostic feature ranking. From these methods, we find that the  signal to noise of the folded profile to be the best feature.  We find that all the ML algorithms report 
 FPRs about an order of magnitude lower than  the corresponding FPRs  obtained in ~\citet{Morello}, for the same recall value.

\end{abstract}

\end{frontmatter}

\section{INTRODUCTION}
\label{i}
\par 
Ever since the accidental discovery of `The Little Green Man'~\citep{Hewish} fifty years ago, a massive amount of work force and computing resources have been invested in exploring the observable Universe to detect radio pulsars.  Pulsars are highly magnetized rotating neutron stars, with misaligned magnetic and rotation axes, which emit pulsed radio emission.  Their radio signals are observable when the axis of their  emission cone is directed along the line of sight to the observer. Subsequently, they have also been observed throughout the electromagnetic spectrum.   An updated review of the various observational properties of  pulsars and other kinds of neutron stars  can be  found in ~\citet{Kaspi}. 
\par
Pulsars  have provided a remarkable laboratory for tests and applications of nearly all branches of physics,  from condensed matter physics to quantum chromodynamics,  and also on a  wide  variety of topics in astrophysics spanning stellar evolution, interstellar medium, cosmology etc.~\citep{Blandford,Ransom}.  They can also be used to provide insights into the nature and distribution  of dark matter~\citep{Afshordi,Desai}. Pulsar observations led to the first confirmed discovery of  extrasolar planets~\citep{Frail}, provided the first indirect evidence for gravitational waves~\citep{Taylor},  and could also provide evidence for the direct detection of gravitational waves in the nHz regime~\citep{Detweiler}. Therefore, it is imperative to discover new pulsars in order to harness  their tremendous physics potential.
\par As of April  2017, there are 2613 known pulsars in the ATNF catalog~\citep{ATNF}. However, the total observable pulsar population in our galaxy has been estimated to be  from $20,000$ to $10^5$~\citep{Yusifov,Faucher,Johnston}. If these estimates are correct, $\sim 90\%$ of the Galactic pulsar population is yet  to be discovered.

Pulsar searches from modern radio surveys involve sifting through candidates detected by pulsar search pipelines, consisting of either periodicity or single-pulse searches. These pulsar search algorithms are often computationally very expensive (although improvements continue to be made to their speed and sensitivity e.g.,~\citealt{Smith,Cameron}). The output of these pipelines yield millions of candidates, out  of which a small fraction consists of pulsars 
and the remaining candidates arise from radio frequency interference (RFI) or other sources of noise~\citep{Keith}. Many of these candidates  are visually inspected and manually vetted by domain experts.  For current generation pulsar surveys, it takes about 1-300 seconds to vet each candidate~\citep{Eatough}.
Therefore, it would require up to 80,000 person hours  to visually vet the million or so candidates. Such a  manual visual classification  of the pulsar candidates   becomes intractable during the SKA era, where we expect to discover around 20,000 new pulsars~\citep{Kramer}. Even though the sifting of real pulsar signals from noise can be facilitated  with graphical utilities such as {\tt JREAPER}~\citep{Keith09}, these have limitations and one is prone to make mistakes~\citep{Bates,Eatough}.  Therefore, to  maximize the detection of pulsars in the SKA era,  the computational costs during  all  the  steps of the pulsar search pipeline should be reduced and human intervention should be minimized at every step. An important step in this process would be to  automate the filtering of pulsar candidates obtained from pulsar search pipelines as much as possible.

Therefore, for the autonomous identification of true signals from noise,  the radio pulsar community has resorted to machine learning to solve this  problem~\citep{Eatough,Bates,Lyon13,Lyon14,Morello,Lyon,Zhu,Wagstaff,Devine}. 
In most of these papers, the machine learning algorithm used is some variant of an Artificial Neural Network (ANN). Supervised machine learning using an ANN was first used in the pulsar community by ~\citet{Eatough} to process 16 million pulsar  candidates obtained  by reprocessing data from the Parkes multi-beam survey.~\citet{Bates} also used an ANN  in the data-processing pipeline for the High Time Resolution Universe (HTRU) mid-latitude  survey. They were able to reject 99\% of the noise candidates and detect 85\% of the pulsars through a blind analysis. 
~\citet{Zhu} used a combination of three different supervised  algorithms, namely an ANN, Support Vector Machine (SVM), and Convolution Neural Nets in their image recognition based post-processing pipeline dubbed PICS (Pulsar Image-based Classification System) AI. PICS was trained with data from the Pulsar Arecibo L-band Feed Array (PALFA) survey and validated with data from the Green Bank North Celestial Cap survey (GBNCC). From the validation set, PALFA was able to rank 100\% of the pulsars in the top 1\% of all candidates, while 80\% were ranked higher than any noise or interference events. \citet{Lyon13} studied the performance of various stream classifiers, such as very fast decision trees  on pulsar data from the HTRU survey. They demonstrated the susceptibility of the  pulsar data  to the imbalanced learning problem and how the imbalance severely reduces pulsar recall. Thereafter,~\citet{Lyon14} presented a new classification algorithm for imbalanced data streams using Hellinger distance measure, which they applied to pulsar data from the HTRU survey. They were able to demonstrate that the algorithm can effectively improve minority class recall rates on imbalanced data.
~\citet{Morello} used neural networks in a pulsar ranking pipeline dubbed Straightforward Pulsar Identification using Neural Networks (SPINN).  This  pipeline was able to  identify all the pulsars in the HTRU-S survey with a false positive rate of 0.64\% and also helped reduce the amount of candidates to scan by up to four orders of magnitude. In this work, we apply multiple machine learning algorithms to the same dataset, as the candidates used for
the cross-validation of the SPINN pipeline were made publicly available~\footnote{\url{http://astronomy.swin.edu.au/~vmorello/}}.~\citet{Lyon} presented a new method for on-line filtering of pulsars from noise candidates using a tree-based machine learning classifier called Gaussian Hellinger Very Fast Decision Tree. This algorithm was able to process millions of candidates in seconds and had recall rates of close to 98\%, when applied to data from the HTRU-1 and LOTAAS surveys with recall rates $>90\%$ and 
false positive rates  $<0.5\%$. It also helped discover about 20 new pulsars from the LOTAAS survey~\citep{Lyon}.~\citet{Wagstaff} incorporated a machine learning based classifier, based on  random forests in their  radio transient detection pipeline, called V-FASTR. This was able to  automatically filter out known event types 
(pulsars and artifacts) with an accuracy of 98.6\% and  achieved a 99-100\% accuracy on newly collected data.
Most recently,~\citet{Devine} used six different machine learning algorithms  to classify dispersed pulsar groups in the second stage of  their single-pulse search pipeline, which is also sensitive to other ``siblings'' of radio pulsars, such as 
Rotating Radio Transients (RRATs)~\citep{RRAT} and Fast Radio Bursts (FRBs) \citep{FRB} 
These six  algorithms included an ANN~\citep{bprml}, SVM \citep{SVM},   direct rule learner \citep{Cohen95}, standard tree rule learner \citep{Quinlan}, hybrid rule-and-tree learner \citep{Frank}, and  Random Forests \citep{RF}. 
Their benchmark dataset consisted of over 300 pulsar signals and about 9,600 noise candidates using observations from the Green Bank Telescope. They found that multiclass ensemble tree learners were the most efficient.
Therefore, in most of  the above papers, some variant of an ANN has been used for the post-processing of candidates from the pulsar search pipelines. 
Here, we start out with an ANN and showcase other algorithms which perform as good as or better than an ANN.

The outline of this paper is as follows. In Section~\ref{sec:ml}, we provide an abridged pedagogic introduction to machine learning, including the definition of  various quality scores. The pulsar dataset from the HTRU-S survey along with the features  used for training are described in Section~\ref{pulsardataset}. In Section~\ref{ml}, we list  the various machine learning algorithms applied to the pulsar dataset. Our implementation of ANN, Adaboost, Gradient Boosting Classifier, and XGBoost  are described in  Sections~\ref{sec:ann},~\ref{sec:adaboost},~\ref{sec:gbc}, and ~\ref{sec:xgb} respectively. We discuss the feature selection procedure  in Section~\ref{fmrmr}.  We report the results of our machine learning algorithms in Section~\ref{sec:res}. We then conclude in Section~\ref{sec:concl}.

\section{SUPERVISED MACHINE LEARNING TERMINOLOGY}
\label{sec:ml}
There are two types of supervised machine learning problems, namely  \emph{classification} and \emph{regression}. 
Given a collection of samples, where each sample uniquely belongs to a group (also known as it's class), classification is the process  of putting samples into their respective classes as accurately as possible. Each sample is described via a collection of variables called `features', which in some way is representative of the characteristics of each class. If for a given collection of samples, its true class labels (used to denote class membership) are also known, it is possible to build a mathematical model, which maps  the `features' (input) of a sample to the predicted class label (i.e., the class to which a sample belongs to) with high accuracy.
A collection of samples for which both the feature data and class labels are known, is referred to  as ``labeled data''. Labeled data is usually bifurcated  into two distinct sets: i) a `training set' used to `learn' a function that maps a sample's features to a label, and ii) a test set used to assess the classification performance.
A special case of classification involving only two classes is called \emph{binary classification}. The separation of real pulsar signals from noise is an example of \emph{binary classification}. The relationship between the features and  their class variables is determined using a training
set, which is a subset of the data for which we know both the features and the class labels. 

 It is a commonplace to use cross-validation to split randomly the available  data into training and testing folds (independent samples) for analysis, to avoid problems associated with over-fitting or under-fitting, which leads to sub-optimal performance in practice. Cross-validation is used to build and test classification models iteratively, yielding an indication of aggregate data set performance (i.e. \rthis{test on fold 1, train on remaining folds, then test on fold 2 and so on, aggregate results}). Cross-validation is thus important for checking that the models we build are capable of generalizing beyond small training sets. The performance on the test set is used as the criterion to select the best algorithm.

More details on supervised machine learning techniques can be found in various monographs~\citep{bprml,Mitchell} and their astronomical applications are reviewed in \citep{Brunner,Cavuoti}.

\subsection{Evaluation measures } 
\label{ss:metric}
\par In order to comparatively rank the performance of various machine learning algorithms, we calculate various performance scores for each algorithm, which we define in this section.  We have used eight different scores\footnote{Score is a floating point number, which quantifies the performance of an algorithm} to quantify the model performance, viz. recall, precision, accuracy, $f_1$ score, log loss, G-Mean, Area under Receiver Operating Characteristics (AuROC) and False Positive Rate (FPR).  

In addition to these eight scores,  we also evaluate the  confusion matrix for every algorithm. A confusion matrix is a $c \times c$ matrix, where $c$ indicates the number of class labels. Each element ($C_{jk}$) of a confusion matrix 
is the number of observations labeled by the classifier as $j$,  but which belong to class $k$. 
For our binary classification problem, the target class is considered to be positive if the candidate being classified is a pulsar and assigned the label of one. If the target class is a noise or RFI candidate, the candidate is assigned a negative class with label equal to 0 or -1. In the case of binary classification problem (our case), each element of the confusion matrix corresponds to TP, TN, FP, and FN, which are defined in Table~\ref{tab:defcon}. In a multi-class classification setting, such a correspondence is absent. 
\begin{table}
\centering
\begin{tabular}{ccc}
\toprule
& \textbf{POSITIVE} & \textbf{NEGATIVE} \\ \midrule
\textbf{TRUE} & TP & TN \\ 
\textbf{FALSE} & FP & FN \\ \bottomrule 
\end{tabular}
\caption{Confusion matrix for a binary classification problem. The rows correspond to the ground truth labels and the columns correspond to the labels as assigned by the classifier. Definitions of the symbols used here can be found in Section~~\ref{ss:metric}}
\label{tab:defcon}
\end{table}
\begin{enumerate}
\item \textbf{TP}: True Positive. Those observations belonging to the TRUE class, which  are correctly labeled by the classifier as positive. 
\item \textbf{TN}: True Negative. Those observations belonging to the FALSE class, which  are correctly labeled by the classifier as negative.
\item \textbf{FP}: False Positive. Those observations belonging to the FALSE class, which are incorrectly labeled by the classifier as positive. Hence, the name False Positive. This type of error is also called   Type-I error.
\item \textbf{FN}: False Negative. Those observations belonging to the TRUE class, which are incorrectly labeled by the classifier as negative. This is also referred  to  as Type-II error. 
\end{enumerate}
Now, we define each of the  eight scores used to evaluate the algorithms used in this paper. 
\begin{description}
\item[Recall] (also known as Sensitivity)  is defined as, 
\begin{equation}
\frac{TP}{TP+FN}.
\end{equation}

A pristine machine learning classifier will have a recall value of   $1$ and the  minimum value  for an egregious classifier  would be  $0$. A Recall of one implies that all positively classified observations belong to the TRUE class. This score is given prominence, when misclassifying a TRUE class observation has a higher cost than misclassifying a FALSE class observation, or  in other words when Type-II error has a higher penalty than Type-I error.  This is the case we need to consider for the pulsar separation problem discussed in this work.
\item[Precision]  is defined as, 
\begin{equation}
\frac{TP}{TP+FP}.
\end{equation}
Precision is the ratio of TRUE class observations labeled correctly to the  total number of  observations, which were labeled as positive  (irrespective of whether they are true positives or not). Similar to recall, even here the best value is $1$ and worst value is $0$. Higher the value of this score, better is the classifier in ascertaining the TRUE class observations.

\item[Accuracy]  is the most popular score in any classification scenario, where the class distributions  are equally balanced, and is  defined as,
\begin{equation}
\frac{TP+TN}{TP+TN+FP+FN}.
\end{equation}
This score has the value of  $1$ as the best value and $0$ as the worst. An accuracy of $1$ implies that the classifier isn't making any error whatsoever. 
\item[$f_1$ score] also known as F-measure or balanced F-score~\citep{bprml}  is defined as, 
\begin{equation}
 \frac{2 \times \rm{Precision}\times \rm{Recall}}{\rm{Precision} + \rm{Recall}}. 
\end{equation} 
 In other words, $f_1$ score is the harmonic mean of precision and recall. It provides an alternate measure of accuracy and has the best value at $1$ and worst value at $0$. \rthis{We note that precision, recall, and accuracy range between values of 0 and 1, where  a value of 1 indicates the absence of errors for that score}. 

\item[Log Loss]  also known as logistic loss or cross-entropy loss  is defined as,
\begin{equation}
\rm{log loss} = - \frac{1}{N} \sum_{i=1}^{N}(y_i\log(p_i) + (1-y_i)\log(1-p_i)).
\label{eq:logloss}
\end{equation}
In Eqn.~\ref{eq:logloss}, $y_i$ represent the  class labels and $p_i$ are the corresponding probabilities. Unlike the other scores discussed above, the range of values of logloss falls between $[0,\infty)$, where a score of $0$ is achieved in the case of a perfect classifier. Interested readers are advised to refer to ~\citet{bprml} for more details. 


\item[G-Mean] is the geometric mean of recall and specificity. It is defined as, 
\begin{equation}
\sqrt{\rm{Recall} \times \rm{Specificity}}.
\end{equation}
Specificity~\citep{spec_roc} is a measure of how well the classifier is able to label a  negative class. It is analogous to recall, which is a measure of how well the classifier is able to label a positive class. The best and worst values of  G-Mean score  are $1$ and $0$ respectively. This score has been given prominence in unbalanced datasets~\citep{gmean_he}.

\item[Area under the Receiver Operating Characteristics] AuROC \citep{auroc} is a score, whose best value is $1$ and worst is $0.5$. Receiver Operating Characteristics (ROC) is a plot showing the  true positive rate versus false positive rate for a classifier. AuROC is the area under this curve, which is computed using the trapezoidal rule. A perfect classifier has an area of $1$. AuROC actually is an area and hence should have values less than $0.5$ as well. However, we can always swap our definitions of what we call `positive' and `negative' classes to get an AuROC score of greater than $0.5$. We get an AuROC score of $0.5$ for a completely random classification.

\item[False Positive Rate] FPR is defined as the ratio of false positives to all the class observations that belong to the FALSE class in the dataset. 
It is defined as,
\begin{equation}
\frac{FP}{TN + FP}. 
\end{equation}
Additionally, since our problem is a binary classification problem, it is also equal to $1 - {\rm specificitiy}$. 
\end{description} 
\section{PULSAR DATASET and SMOTE}
\label{pulsardataset}
\subsection{Pulsar Detection}
We  briefly outline  the steps involved in the detection of  pulsars from raw radio data  using periodicity searches.  More details on each of these steps can be found in ~\citet{Lorimer} and ~\citet{Lyon} and references therein.
The first major step in the pulsar search pipeline  after data digitization, channelization, RFI excision and `clipping'~\citep{Hogden}, involves de-dispersion, which corrects for the frequency dependent delay the incoming signals experience due to the presence of free electrons in the interstellar medium. The amount of dispersion is governed by the integrated electron density along the line of sight, also  known as the dispersion measure (DM). However, the DM is generally not known in advance. Therefore, we must search over a large range of possible DMs, to find the optimum DM that achieves the highest S/N for a suspect detection.
A brute force search algorithm for pulsars involves a grid search in DM and acceleration space. For each putative DM and acceleration, periodic signals are searched for  using FFT techniques and potential pulsar candidates are then selected from these searches. These candidates are then manually inspected to identify the pulsar-like sources, which later can be analyzed.

\subsection{HTRU Survey}
\label{datum}
The dataset we have used in this manuscript for testing various  algorithms is obtained through the reprocessing of data  from the Southern High Time Resolution   mid-latitude survey~\citep{Keith} by the {\tt PEASOUP} pipeline. Briefly, the  Southern HTRU Latitude survey (HTRU-S) is a survey of the southern sky ($30^{\circ}>l>-120^{\circ}$; $|b|<15^{\circ}$),   carried out using the 13-beam  receiver on the Parkes Radio Telescope to detect radio pulsars and also short-duration radio transients.  A similar survey of  the Northern skies has also been conducted using the Effelsberg radio telescope with a seven-beam receiver~\citep{HTRUN}. More details of the HTRU-S survey strategy and specifications can be found in ~\citet{Keith}. This data has been reprocessed  using a GPU-based {\tt PEASOUP} pipeline with acceleration searches  turned on~\citep{Morello} (hereafter M14). The pipeline searched for dispersion measures between 0 and  400 $\mathrm{cm^{-3}}$ pc and accelerations between $-50$ and $50$ $\mathrm{m/s{^2}}$. 
M14 manually processed their candidate set and constructed a dataset for machine learning purposes. This dataset consisted of  1200 real pulsars and to this,  90,000 non-pulsar observations (from RFI and other sources of noise) were added.
The pulsar candidates encompass a diverse range of periods, duty cycles, and signal to noise ratios. It is unlikely that  there are  undiscovered pulsars in the set of noise candidates.  This dataset (with the pulsar and noise candidates tagged for each event), was made publicly available  by M14   for the development and testing of  various machine learning algorithms by the  wider scientific community. Our goal is to find a machine learning solution with the  maximum efficiency in detecting pulsars and minimum contamination from noise candidates. In machine learning parlance as discussed earlier, we are tackling a binary classification problem.

\par
\begin{figure*}
	\includegraphics[width=0.88\textwidth,keepaspectratio]{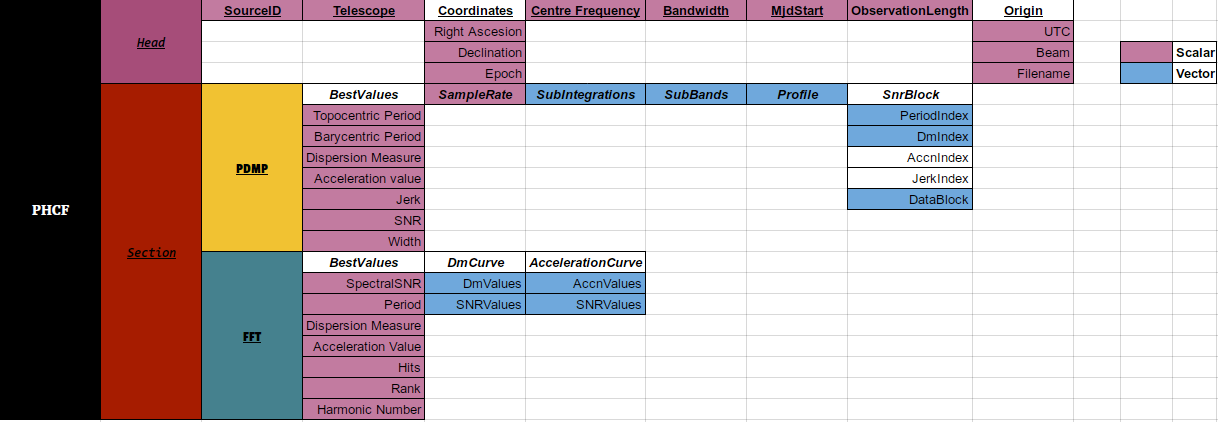}
    \label{fig:schema}
    \caption{Schema of the XML tree containing details of the pulsar dataset from the HTRU survey, described in detail in M14 and made publicly available. Each candidate file has the above schema and structure. }
\end{figure*}

 Each pulsar candidate is described in a PulsarHunter Candidate XML file, having the extension {\tt .phcx}. See Figure~\ref{fig:schema} to view the schema of the xml tree. 
\par For all the machine learning algorithms herein, we have used the same feature set as defined by M14. They are listed below, and henceforth we shall denote each feature with a corresponding numerical index~(called \emph{feature index}, $f_{ID}$), using $0$-based indexing. More details on each of these features can be found in ~\citet{Lorimer}. We have also included a Python script file in the Github repository, by the name of \texttt{ExtractFeatures}, which can be used to extract the following features from the PHCX files. 

\begin{enumerate}[label=\arabic*.]
\setcounter{enumi}{-1}
\item S/N Ratio of the folded profile:
S/N ratio is a measure of the signal significance and is defined for a contiguous pulse window $W$ as~\citep{Lorimer,Morello},
\begin{equation}
S/N = \frac{1}{\sigma\sqrt{w}}\sum_{s_i \in W} (s_i - \bar{b}), 
\end{equation}
\noindent where $s_i$ is the amplitude of the $i$-th bin of the folded profile, $w$ is the width of the pulse region in each bin, $\bar{b}$ and $\sigma$ represent the mean value and the standard deviation of the folded profile in the off-source region. Because of the large dynamic range of the S/N ratio, its logarithm is used as a feature.
\item Intrinsic equivalent duty cycle of the pulse profile:  This is an approximate proxy for the total duration of the signal and is defined as the the ratio of the pulse width to its spin period. However, for millisecond pulsars one needs to correct for the large duty cycles due to dispersive smearing. The intrinsic equivalent duty cycle of a candidate is defined as,
\begin{equation}
D_{eq} = \frac{w_{eq} - \Delta \tau}{P},
\end{equation}
where $P$  is the period of the candidate and $\Delta \tau$ is the approximate dispersive smearing time and is defined in M14.
\item Ratio between barycentric period and dispersion measure~\emph{(log scale)}: This is the logarithm of the ratio of the period and the dispersion measure.
\item Validity of dispersion measure: This is a figure of merit to bifurcate the dispersion measures of pulsars and RFI signals and is defined as,
\begin{equation}
V_{DM} = \tanh (DM - DM_{min}),
\end{equation}
\noindent where $DM$ is the dispersion measure of the candidate and $DM_{min}=2$ for HTRU data. This feature is used to reject non-pulsar candidates below a certain threshold.

\item Persistence of signal in the time domain: This is based on the \emph{ansatz} that a true pulsar will  be consistently visible in the time domain for most of the observation and is defined following~\citealt{kjlee},
\begin{equation}
\chi(s) = \begin{cases} 
					1-\exp(-s/b) & \text{if}\  s \geq 0 \\ 
				    s/b &  \text{if}\   s <  0,  
           \end{cases} 
\end{equation}
\noindent where $s$ is the S/N ratio of a candidate in a sub-integration, and $b$ is the benchmark S/N ratio, which is a user-defined parameter and the choice adopted by M14 is  $b = \frac{2 S_{min}}{\sqrt{n_{sub}}}$. 
\item RMS distance between the folded profile and the sub-integrations. This is a measure of the variability of the pulse shape from the observations. The folded profile as well as every sub-integration is normalized to values between 0 and  1.  The root mean square distance between the folded profile of candidate and each of its sub-integrations is defined as,
\begin{equation}
D_{RMS} = \sqrt{\frac{1}{wn_{sub}}\sum_{i \in W}\sum_j(p_i-s_{ij})^2},
\end{equation}
where $p_i$ is the value of the $i$-th bin of the folded pulse profile, $s_{ij}$ is the value of $i$-th bin of the $j$-the sub-integration.
\end{enumerate}
 All of the above features  were scaled to have zero mean and unit variance, before training.
There are only $1,192$ pulsar observations in the dataset, whereas there are $87,680$ non-pulsar observations. For completeness, we report the performance of all the ML algorithms on both the original dataset  and on an artificially inflated version of the dataset using Synthetic Minority Over-sampling Technique (SMOTE) (described in the next subsection), which accounts for the mismatch between the ratio of the pulsar to non-pulsar candidates. The motivation behind doing so is to compare how these algorithms perform on highly skewed datasets and also not to bias the results from the ML algorithms.

\subsection{SMOTE}
\label{ss:smote}
Synthetic Minority Over-sampling Technique, popularly known as SMOTE~\citep{smote}, is a very well known technique employed in high class imbalance problems. We note that high class imbalance occurs when the class distribution in a classification problem is skewed toward a particular category (either in the training data, or the real-world data). A ratio of 1:10 is generally considered as imbalanced, and 1:100 or greater is considered as  highly imbalanced~\citep{gmean_he}. Our dataset has a ratio of 1:73 making it a high class imbalance problem. 
The crux of this technique involves manufacturing artificial vectors in the feature space to create synthetic datasets, so as to make the ratio more uniform. 
This generation of mock data  is done until the ratio of the minority  to the majority class population reaches a user specified threshold, which is passed as an argument to the SMOTE algorithm.  For our implementation of SMOTE, k-Nearest Neighbors (kNN,~\citealt{knn_ref}, Chapter 13 of~\citealt{eslii}) is employed in the process of creation of synthetic feature vectors. For every feature vector belonging to the minority class, it's $k$ nearest neighbors are determined, from which a random neighbor is chosen and 
in-between this randomly chosen neighbor and initial feature vector, synthetic feature vectors are interpolated randomly. This is a bird's-eye view of the algorithm. Interested readers should refer to~\citet{smote} for more details. 

SMOTE has been previously used for the classification of unknown point sources from the Fermi-LAT catalog~\citep{fermi} and
classification of variable stars from Kepler~\citep{Borne}. Two years ago, it has also been used in a pulsar search pipeline to   address the class imbalance problems~\citep{Devine}.

We run all our machine-learning algorithms on two datasets, the original one, on which no SMOTE has been applied, which we refer to as ``\textit{non-SMOTE}'' applied data,  and another one with SMOTE applied,  which we refer to as ``\textit{SMOTE}''. 
We use SMOTE to artificially balance the dataset, i.e. to have a 1:1 ratio of pulsar to non-pulsar samples.

\par Our approach of addressing the high class imbalance problem is  complementary to what M14 have done. M14 performed oversampling of the pulsar-only dataset, until  they  have four non-pulsar (negative) samples for each pulsar (positive) sample in the training dataset. On the contrary, we employed SMOTE to artificially balance the dataset. We have used the Python module called \texttt{imbalanced-learn} to apply the SMOTE technique to our datasets.

\section{MACHINE LEARNING METHODS}
\label{ml}
In this work, we have applied four different algorithms to this pulsar classification problem: 
\begin{enumerate}
\item Artificial Neural Network (ANN)
\item Adaboost
\item Gradient Boosting Classifier (GBC)
\item eXtreme Gradient Boosting (XGBoost)
\end{enumerate}

\par To the best of our knowledge, among these four algorithms, Adaboost, GBC, and XGBoost have never been used before in the core pulsar search pipelines or during the post-processing of pulsar candidates. Although, we are only reporting the aforementioned techniques, we also tried the  Na\"ive Bayes algorithm. However, the performance of the  Na\"ive Bayes was unsatisfactory. 
The core premise  of  the Na\"ive Bayes algorithm is conditional independence between the features for a given target label. We find that there is  significant amount of correlation between the features. Hence,  Na\"ive Bayes fails on this dataset~\citep{optNB}. Therefore, we do not report the results from the Na\"ive Bayes algorithm. We note that this agrees with the conclusions from~\citet{Devine} and ~\citet{smith13}, who also found that Na\"ive Bayes performs poorly on pulsar data. However,~\citet{Lyon14} and \citet{Lyon} observed reasonable performance with this algorithm using a different feature set on the HTRU-1, HTRU-2, and LOTAAS-1 datasets.

 We now describe in detail our implementation of the four algorithms and present their classification results. All four algorithms have been implemented in Python using the {\tt scikit-learn} module~\citep{scikit-learn}. The implementation code, along with the dataset, is available in the Github repository\footnote{\url{https://github.com/shiningsurya/pulsarml}}. We have  also added the Python code to generate the dataset (\texttt{csv}) file from the  PHCX Candidates files. More details about the code can be found in the Github repository.
In a typical $k$-fold cross-validation (CV) procedure, the entire dataset is split into $k$-folds and the classifier under test is trained on $(k-1)$-folds and assessed on the remaining one fold, which was not used for training. This process is repeated $k$ times so that the classifier under test can be assessed (by computing the scores) on each fold of the dataset. To further reduce the variance in the scores, the complete $k$-fold CV is performed multiple times and the scores are averaged over these iterations. A variant of the CV procedure, where the dataset is judiciously broken into folds such that the ratio of positives to negatives in each fold is preserved as best as possible, is called stratified CV, which we use here. In the non-SMOTE case, the classifier under test is simply trained on the training fold. However, in the case of SMOTE, we employ SMOTE in the CV procedure to artificially balance the training fold (and only the training fold) and on this inflated training fold, the classifier under test is trained. It is crucial to note that in both the cases (SMOTE and non-SMOTE), the classifier is tested on the same test fold. We also  note that the hyper parameters for  all the ML models (cf.~Table~\ref{tab:all_hp}) are the  same in both the non-SMOTE and the SMOTE cases.

\par We carry out $20$ iterations of $5$-fold stratified CV for each of the machine learning algorithms discussed in this work using the procedure discussed above. We therefore get $100$ values for each of the scores, which we aggregate and report.
\begin{table}
\centering
\caption{Hyper parameters of all the  machine learning (ML) models discussed here. We have used the same set of hyper parameters for  both the  non-SMOTE and SMOTE datasets.}
\label{tab:all_hp}
\begin{tabular}{ll}
\toprule
ML & Hyper-parameters                                                                                                          \\ \midrule
\multirow{7}{*}{ANN(MLP)}    & 6:6:4 network \\ 
                             & activation function : Leaky RLeU \\
                             & output layer : softmax \\
                             & learning rule : adam \\
                             & learning rate = 0.001 \\
                             & batch size = 32 \\
                             & number of iterations = 750 \\ \midrule
\multirow{3}{*}{Adaboost}    
& number of estimators = 250 \\  
& learning rate = 0.1 \\
& base estimator: decision stump   \\ \midrule
\multirow{4}{*}{GBC}         & number of estimators = 250 \\
                             & learning rate = 0.1 \\
                             & max depth = 5 \\
                             & minimum number of samples at leaf node = 2 \\     \midrule                                                                  
\multirow{5}{*}{XGBoost}     
& number of estimators = 500 \\
& learning rate = 0.001 \\
& maximum depth = 3 \\
& regularization alpha = 0.0001 \\
& gamma = 0.1                                                              
\\ \bottomrule
\end{tabular}
\end{table}
\subsection{ANN (MLP)}
\label{sec:ann}
\par An Artificial Neural Network (ANN) is a popular tool in the machine learning community. Complete details of an ANN can be found in~\citet{bprml} and \cite{eslii}. Most machine learning algorithms used by the pulsar community are some variant of the ANN. ANNs are also widely used in optical astronomy~\citep{Lahav}, in particular for photometric redshift estimation (eg.~\citealt{BCS}). 

\par We have  used a simple Multi-Layer Perceptron (MLP) similar to M14. The hyper parameters used for training the ANN are specified in Table~\ref{tab:all_hp}. We make use of the {\tt SKNN} Python module to implement this MLP.\footnote{Please note of the dependency issue of SKNN with {\tt theano}.} 
 
More details on our implementation can be found in the  Github repository. The hyper parameters, along with all the connection weights, form a functional mapping from the input features to the output (target variable). The loss function is defined on the predicted target and the true target (which comes from the labeled data).
\par Minimization of this loss function is a high dimensional non-convex problem. Hence, we use a gradient-based method to perform the optimization. The heuristics are as follows: Our choice of hyper parameters and training data yields a specific level of training error. We then compute the gradient and update the connection weights accordingly (in the direction of negative gradient) so as to reduce the training error. Our aim is to find the most optimum set of connection weights that lead to the lowest training error. The most crucial step here is the computation of the connection weights from the gradient, and for this purpose, we employ the learning rate.
Learning rate is defined as the hyper parameter which regulates the magnitude of the distance we want to traverse against the direction of its gradient. 
For  all the gradient-based solutions, determining the optimal learning rate is often the trickiest problem. If we choose too high a value of this hyper parameter, we might never converge to the global minima (maxima) but could  pass right by it, whereas too low a value of the learning rate could possibly increase the convergence time and the possibility of  getting stuck in a local minimum (maximum). At times, finding an optimal parameter becomes so difficult that switching to alternate machine learning solutions may seem more prudent. 
In order to mitigate this issue, we have  used {\tt Adam} (a method for stochastic optimization) as our gradient descent optimization algorithm (learning rule), instead of the default Stochastic Gradient Descent (SGD) in {\tt SKNN}.
In the SGD optimization algorithm, the learning rate is either  a constant or is reduced after  a fixed number of epochs. {\tt Adam}, instead computes the running first and second-order (un-centered) moments of the gradients and dynamically adjusts the learning rate on the fly.
The base learning rate (`original learning rate') is normalized   by dividing by the square-root of the non-centered and bias corrected second-order moment of the gradients, and followed by scaling using  the running bias-corrected first-order moment of the gradient as the scale parameter. It is important to note that these gradients essentially decay exponentially,  since the neural network is converging. There is also an additional fudge factor added to this square-root (second order moment), in order to ensure that the division operation doesn't throw a `divide-by-zero' exception. More details about {\tt Adam} can be found in \citet{adam}. It is also worthwhile to note that Adam draws it's name from ``ADAPTIVE'' moments, since the running moments determine the learning rate.  

\par We note that M14 employed a single hidden layer with eight hidden units and used {\tt $\tanh$} as the output layer activation function. They have trained using the `mini-batch' approach, wherein the weight updates happen after a batch (subset) of the  data has been processed. On the other hand, we have implemented a 6:6:4 MLP and also  used a different learning rule (Adam), activation function (Leaky RLeU~\citealt{leaky_rectify}), and the output layer of {\tt softmax}~\citep{softmax}.
The {\tt softmax} layer ensures that the  final output  lies between $[0,1]$ (analogous to probabilities), contrary to $\tanh$ function, whose range is  $[-1,1]$. 
\subsection{Adaboost}
\label{sec:adaboost}
\par Adaboost, an acronym  for Adaptive Boosting, takes \rthis{a collection (an ensemble)} of `weak learners' and recursively  trains them on copies of the original dataset, all the while focusing on the `difficult' (or outlier) data points \emph{(hence the word `Adaptive')}~\citep{Adaboost}. This, in turn, ensures that such an ensemble performs drastically better on the test data. 
A natural choice of weak learners are simple rules or logic, which are realized as decision trees, also known as algorithmic flowcharts~\citep{cart_trees}.

Those decision trees which are of depth one, are called `decision stumps'~\citep{decision_stump}.
More details about Adaboost can be found in  Section 2 of \citet{Mayr}. Adaboost has been shown to outperform other machine learning algorithms for  many different supervised  classification problems in optical astronomy~\citep{fimp_adaboost,Sevilla,Elorrieta,Acquaviva,Zitlau}. 

\par We have used the same implementation of Adaboost as in ~\citet{fimp_adaboost}, where it was applied to photometric redshift estimation of SDSS DR10 data and shown to be more robust against outliers and provided better matches with spectroscopic redshifts than the ANNs.  We have defined our Adaboost model using two properties (specified in Table~\ref{tab:all_hp}), viz., the number of trees and the learning rate. The former corresponds to the number of weak learners in the ensemble and the latter determines how strongly each weak learner should contribute to the weights. For a large number of estimators, each estimator would have it's own effect on calculating the weights, which would get enhanced strongly. On the contrary, for a smaller number of estimators, the weights assigned may not be sufficient for the model to learn a `tricky' sample (hard to classify).
In other words, there exists a trade-off between the number of estimators and the learning rate. Please note that Adaboost is a meta-classifier~\citep{meta_eml},  which takes {\tt N copies} of the weak learner and trains them on the same feature set but with different weights assigned. These {\tt N copies} correspond to the number of weak-learners. See \cite{opt_ada}, which extols the power of Adaboost on mathematical grounds.

\subsection{Gradient Boosting Classifier}
\label{sec:gbc}

\par  A Gradient Boosting Classifier (GBC) is a form of ensemble classification system. Much like Adaboost, which was discussed in the previous subsection, GBC iteratively adds simple classifiers (decision trees, although not necessarily  of depth one), which are successively  trained on the errors of the predecessor classifier. 
In other words, the very first classifier of the ensemble is trained on the dataset, whereas the second classifier is trained on the errors of the first classifier and added to the first classifier and so on. \rthis{The merging takes place in an aggressive fashion; the successor tree couples with the predecessor tree using a coupling parameter, which is optimized so that the error of the combined system is minimized. Note that, this is formally a linear optimization problem.  }
This hierarchical procedure is also  known as  boosting. See~\citet{gbc11,gbcas} for a thorough understanding of Gradient Boosting Machines. Gradient boosting has also been widely used in astrophysics for photometric classification of Type 1a supernovae~\citep{SN}  and for automated galaxy detection and  classification using the Galaxy Zoo catalog~\citep{gbcas}. 
Another hyper parameter  in GBC is the `learning rate', which governs  how strongly a classifier should be merged with it's predecessor. 
Learning rate lends itself to GBC as a regularization mechanism. 
 In GBC, we employ $4$ hyper parameters: learning rate as defined above, number of estimators (n\_estimators, number of trees in ensemble), maximum depth, and minimum samples per leaf, the last two are actually hyper parameters of the decision tree (member of the ensemble). We note that all the members of the ensemble have the same hyper parameter set which can be found in Table~\ref{tab:all_hp}. We followed the same procedure as discussed at the start of this section. We used the {\tt scikit-learn} implementation of GBC.

\subsection{XGBoost}
\label{sec:xgb}
\par XGBoost~\citep{xgboost}, is an acronym for  eXtreme Gradient Boosting, and  works in the same way as Gradient Boosting but with the addition of  Adaboost-like feature of assigning weights to each sample. 

\par XGBoost is a tree-based model which gained a lot of popularity right since it's inception in the machine learning community in 2016. This model was also the winner of \emph{HEP meets ML} Kaggle challenge~\citep{xgboost_wins} and our problem is no exception to that. In  astrophysics,  XGBoost was recently used for the classification of unknown point sources in the Fermi-LAT catalog~\citep{xgas} and also for studying the stability of planetary systems using physically motivated features with the help of supervised learning~\citep{Tamayo}.

Unlike in GBC, where every member of the ensemble (tree) is built sequentially, XGBoost on the other hand, can parallelize this task and give substantial speed boost (See Sections 4,5 of ~\citet{xgboost} and references therein). Moreover, the regularization techniques, such as L1 and L2, which are used to control over-fitting are available in XGBoost, but not  in Adaboost and GBC. Another key feature of XGBoost is the scalability it offers~\citep{xgb_scale}. This means that it can be efficiently run on  distributed systems and can also work on very large datasets with ease\footnote{Refer to this url for the list of use-cases: \url{https://github.com/dmlc/xgboost/blob/master/demo/README.md\#usecases}}. XGBoost can also be trained using Graphical Processing Units (GPUs, See~\citealt{xgb_cuda}) and can offer extremely high speed boosts.
While our dataset is not as onerous as those on which XGBoost is usually run, ours  is the  first proof of principles application of XGBoost  to the pulsar classification problem.

\par XGBoost is readily available as a Python package, which is used in this work. We use five hyper parameters for  training in XGBoost. We build an ensemble of $500$  estimators with a learning rate of 0.1\% (doubled the number of trees compared with Adaboost, GBC because learning rate was changed to $0.001$ to account for the `learning-rate/number of trees` trade-off~\footnote{For a large number of trees, higher learning rate would make our model over-fit to the data. Hence, a smaller learning rate should go with a larger number of trees so that our model can learn the overall structure and not the intricate nuances (noise) of the signal}). 
Every tree in the ensemble has a maximum depth of $3$ with $L1$ regularization parameter of 0.01\%. We have also used the {\tt gamma} hyper parameter which controls the tree building process by constraining the amount of gain a split should have.
\section{Feature Selection}
\label{fmrmr}

\par  We now delve  into the process of feature selection.  In any supervised ML algorithm, the features that a model learns from largely determine the accuracy of the model. An information-theoretic way to quantify the efficacy of a feature is to compute the  Mutual Information (MI) between a feature and its corresponding class label \citep{MI}. MI is defined as the amount of information conveyed by one random variable through another random variable (Here, we treat the features as some random variables). Mathematically, 
\begin{equation}
MI = \int \int p(x,y) \log \frac{p(x,y)}{p(x)p(y)},
\end{equation}
 where $p$ represents the probability density function (p.d.f), whereas $x$ and $y$ are two continuous random variables. Estimating the entropy from  first principles (by computing the p.d.f from the histogram and carrying out the integration) is a very computationally challenging task. To alleviate this, we use the {\tt sklearn.FeatureSelection}~\footnote{This functionality was added in version 0.19} module, which has functions to compute MI. These functions actually use the estimator which is based on~\citealt{mi_kraskov}.

\par We  use the  minimum Redundancy Maximum Relevance \\ (mRMR) procedure \citep{bigmrmr,MI2,MI} to determine the feature rankings using the non-SMOTE and SMOTE datasets. This procedure is widely used in the machine learning community for feature selection~\citep{featurereview}, as it provides an algorithm-agnostic criterion for selecting the best feature and its computational cost is not onerous. Although, mRMR has never been used 
in electromagnetic astronomy before (to the best of our knowledge), it is routinely used in neutrino astronomy, in particular for the IceCube event selection pipeline to select the neutrino events from the background of cosmic ray muons~\citep{icecube}.
For this procedure, we assign a rank to the feature based on it's score which is defined as:
\begin{equation}
\rm{Score}_i = MI(x_i;y) - \frac{1}{d-1} \sum_{j \neq i} MI(x_i;x_j),
\end{equation}
 where  $x_i$ denotes the $i^{th}$ feature column and $d$ represents  the number of features. This score takes into account the mutual information (MI) between a feature and its class label (or how powerful the feature is) and also penalizes for any correlations between the features. Naturally, higher the score, more conspicuous is the feature, and hence its rank should be higher. MI between a feature and the corresponding class label (relevance) doesn't involve any other feature, and  hence  does not measure the inter-feature relationships. 
 Therefore, by definition, mutual information does not include any correlation between the features. The score is defined in such a way to account for `relevance' and `redundancy'.

\par Tree-based ML methods, discussed in this work, provide `feature importances' out of the box. At the grassroot level of any tree-based ML method, a feature is chosen, which is used to make a split (corresponding to a branch in the decision tree), and the number of times a feature is chosen to split  can be treated as a measure of the importance of the feature. \rthis{However, in practice the maximum reduction in the error when a feature is used to make the split is used to gauge it's feature importance. But, still our understanding is valid because if a feature is frequently used, it has better descriptive power and naturally is assigned a higher rank.} The criteria for selecting a feature to partition the data is related to our problem statement. In a regression setting, the objective is to reduce the variance in the partitions so that every sample in any partition is representative of each other. On the other hand, in a classification setting (such as the problem at hand), only a feature which separates the data in such a way that each partition is homogeneous and doesn't contain any contamination is chosen. However, the motivation that the samples belonging to a particular partition are representative of each other remains the same as in a regression setting.
Naturally, features which are chosen larger number of times are more powerful (in terms of their separation ability) and as a natural outcome, the tree-based methods provide additional diagnostic insights about the data on which they are trained. More details on these aspects of feature selection can be found in  ~\citet{fimp_adaboost} and Chapter 10 of ~\citet{eslii}. 

\par
\begin{figure*}
\centering
\caption{Feature rankings as found by Adaboost, GBC, XGBoost, and mRMR approach in this work. We find that the S/N Ratio of the folded profile ($f_{ID} = 0$) to be the most descriptive feature. $f_{ID} = 1$, $f_{ID} = 2$, $f_{ID} = 3$, $f_{ID} = 4$ and $f_{ID} = 5$, correspond to the Intrinsic equivalent duty cycle of the pulse profile, ratio of barycentric period and DM, validity of DM, RMS distance between folded profile, persistence of signal in the time domain, and the sub-integrations respectively. Refer to last paragraph of Sect.~\ref{fmrmr} for more details about this plot.}
\label{fig:feature_ranking}
\includegraphics[width=\textwidth,keepaspectratio]{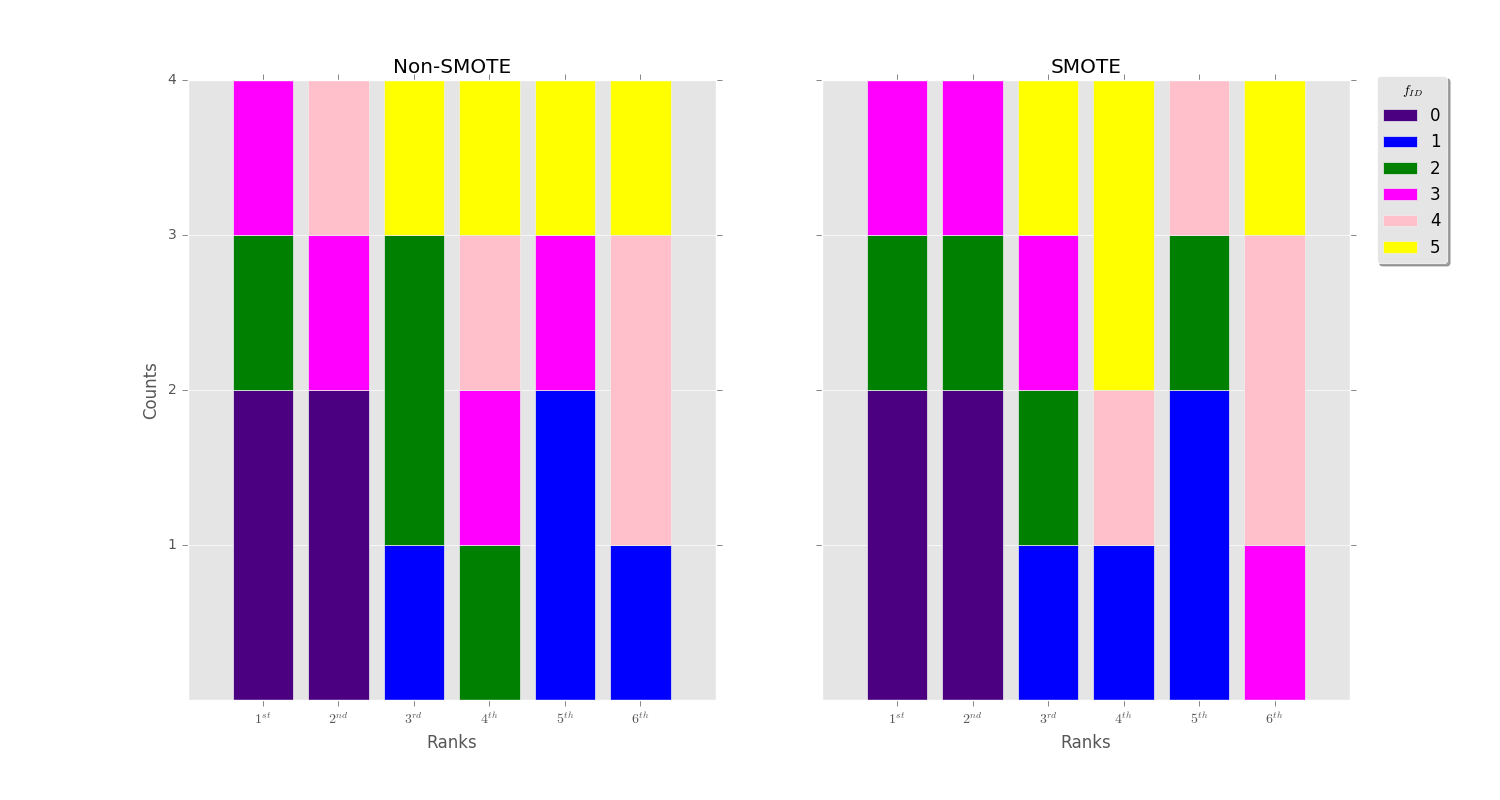}
\end{figure*}

 \bthis{To interpret the feature rankings coherently, we have presented the data using a stacked bar plot in Fig.~\ref{fig:feature_ranking}. For both the non-SMOTE and SMOTE cases, we have four different rankings based on Adaboost, GBC, XGBoost and mRMR. For every rank, we plot the number of times a given feature was assigned that rank in a stacked bar plot, with each feature having a unique color code. For instance, $f_{ID} = 0$ ranks in first and second places twice in both the non-SMOTE and SMOTE cases, which means, that this feature has the most descriptive power. This result is in agreement with~\citet{Lyon}. Lastly, we note that the distribution of features with respect to the  ranks is similar in both the non-SMOTE and SMOTE cases, which implies that our technique of artificially inflating the dataset has not significantly altered the descriptive power of the features in the dataset.}

\section{Results}
\label{sec:res}
\begin{table*}
\centering
\caption{Averaged confusion matrices of all the algorithms  used in this paper. Elements of the confusion matrix are counts, which are integers. However, we have   averaged over the $20 \times 5$-fold Stratified CV, and hence, the floating point  numbers are reported for each of the scores.  Refer to Section~\ref{sec:ml} for details about the  various definitions used here. \rthis{The error bars for all the elements of all the averaged confusion matrices are of $\mathcal{O}(0.01)$ and hence not reported in the table explicitly.} We note that T, F stand for True and False respectively, whereas N, P stand for negative and positive respectively. Refer to Table~\ref{tab:defcon} to understand the confusion matrix. The shaded values are for the SMOTE dataset, whereas non-shaded values are for the non-SMOTE ones.}
\label{tab:cfm}
\begin{tabular}{ccccccccc}
\toprule
ML Model & \multicolumn{2}{c}{ANN(MLP)} & \multicolumn{2}{c}{Adaboost} & \multicolumn{2}{c}{GBC} & \multicolumn{2}{c}{XGBoost} \\ 
\midrule
  & N       & P     & N       & P     & N       & P      & N       & P     \\
F & 1.4     & 10.4  & 1.2     & 3.3   & 1.1     & 2.9    & 0.8     & 5.2   \\
T & 17525.6 & 237.0 & 17532.8 & 237.2 & 17533.0 & 237.3 & 17530.8 & 237.7 \\
\rowcolor{black!20} F & 0.4     & 9.6   & 0.5     & 5.6   & 1.3     & 3.5    & 0.6     & 6.3   \\
\rowcolor{black!20} T & 17526.4 & 238.0 & 17530.4 & 237.9 & 17532.5 & 237.0  & 17529.7 & 237.8 \\
 \bottomrule
\end{tabular}
\end{table*}
\begin{table*}
\centering
\caption{Various scores averaged over $20 \times 5$-fold stratified CV. Refer to Section~\ref{sec:ml} for  definitions of the scores used here, and for the methodology applied to compute these scores. Error bars are negligible (variation in the third decimal digit only) for all the scores, and hence not reported here.  The shaded values are for the SMOTE dataset, whereas the non-shaded values are for the non-SMOTE ones.}
\label{tab:all_scores}
\begin{tabular}{lcccccccc}
\toprule
ML Model & Recall & Precision & Accuracy & $f_1$ & log-loss & G-Mean & AuROC & FPR[\%]\\ 
\midrule
 ANN (MLP)                 & 0.990                             & 0.964                               & 0.999                    & 0.977                               & 0.021                                                      & 0.995                             & 0.994    & 0.059                           \\ 
\rowcolor{black!20} ANN (MLP)                 & 0.998                             & 0.961                            & 0.999                      & 0.979                                     & 0.019                                                       & 0.998                             & 0.994    &  0.055                         \\  
Adaboost                 & 0.995                             & 0.986                                & 0.999                                  & 0.990                         & 0.008                                                        & 0.997                             & 0.994   &  0.019                          \\ 
\rowcolor{black!20}                            Adaboost                 & 0.997                             & 0.976                                & 0.999                                  & 0.986                         & 0.012                                                      & 0.998                             & 0.994                   &  0.032          \\ 
GBC                      & 0.995                             & 0.987                                & 0.999                                  & 0.991                        & 0.008                                                       & 0.997                             & 0.994    & 0.017                           \\  
 \rowcolor{black!20}                     GBC                      & 0.995                             & 0.986                                & 0.999                                  & 0.990                         & 0.008                                                   & 0.997                             & 0.993                            & 0.020 \\ 
                                     XGBoost                  & 0.996                            & 0.978                               & 0.999                                  & 0.987                        & 0.012                                                      & 0.998                             & 0.994                            & 0.030 \\ 
\rowcolor{black!20}                   XGBoost                  & 0.997                             & 0.974                               & 0.999                                  & 0.985                         & 0.013                                                     & 0.998                             & 0.994                            & 0.036\\ \bottomrule
\end{tabular}
\end{table*}

\par  We now report the salient results from applications of all the four ML methods.

The confusion matrices of all the ML algorithms, on both the SMOTE and non-SMOTE
datasets are shown in Table~\ref{tab:cfm}.
A tabular summary of  the various quality metrics can be found in Table~\ref{tab:all_scores}. We make the following observations:
\begin{itemize}
\item All the algorithms give approximately the same values for the confusion matrices for both the datasets. The number  of false pulsars is largest for the ANN (MLP).
\item The recall, accuracy, $f_1$ score, G-mean, and AuROC perform about the same for all the four  algorithms, for both the
SMOTE and non-SMOTE datasets.
\item Precision varies across the datasets, depending on the ML algorithm applied.  In case of Adaboost, the difference between non-SMOTE and SMOTE precision is the largest, and is smallest in the case of GBC, which is surprising, since both these ML algorithms are tree-based. 
\item The log-loss for the ANN (MLP) using the non-SMOTE datasets is about an order of magnitude worse than that for the other algorithms. However, they are comparable, when considering the SMOTE datasets.
\item For all the algorithms, we have been able to achieve a recall value of close to 100\%, but with a FPR about an order of magnitude lower than the value of 0.64\%, obtained in M14. Among the four algorithms, ANN (MLP) shows the largest FPRs for both the non-SMOTE and SMOTE cases, compared to the others.
\end{itemize}

\section{Discussions and Conclusions}
\label{sec:concl}
\par We have run  multiple supervised machine learning algorithms  to evaluate  the efficacy and robustness of the separation of true pulsar signals from noise and RFI candidates, using data from the HTRU-S  survey, a part of which has been made publicly available by M14. We have reported  our implementation and results
from four of these algorithms, namely an ANN Multi Layer Perceptron (ANN (MLP)), Adaboost, Gradient Boosting Classifier (GBC), and XGBoost,  all of which report high accuracy and G-mean.  The last three algorithms have never  been previously applied either  in the core pulsar searches, or in any post-processing pipeline used to separate the real pulsars from noise candidates.

In addition to the feature ranking based on importances, as returned by the tree-based machine learning methods (Adaboost, GBC and XGBoost), we have also implemented feature selection using the Minimum Redundancy and Maximum Relevance Feature Selection (mRMR) approach. This is the first application of mRMR in electromagnetic astronomy.
To alleviate the high class imbalance (high non-pulsars candidates to pulsars candidates), we have used the Synthetic Minority Over-sampling Technique, also known as SMOTE, to artificially balance the datasets.  
For the same set of hyper parameters, each ML method was trained on the original (non-SMOTE) as well as the  SMOTE applied dataset in a $20 \times 5$-fold Stratified CV fashion. We evaluated the various scores for all four of these algorithms and reported the average scores, which can be found in Table~\ref{tab:all_scores}. 

\par Compared to M14, which achieved a FPR of $0.64\%$ at a recall of $100\%$, our ANN (MLP) implementation achieved FPRs of  $0.059\%$, $0.055\%$ for the values of recall of $0.990$, $0.998$ in non-SMOTE and SMOTE cases respectively, . However, M14 performed over-sampling to alleviate the high-class imbalance, whereas, we  did not address  it for the non-SMOTE case, and applied SMOTE to balance the dataset in the other case artificially.   Our alternate implementation of  ANN (MLP) as well  as the tree-based ML methods  resulted in FPRs about  an order of magnitude lower than that of M14 for  both  the non-SMOTE and SMOTE cases.

\par An ideal classifier must be able to classify both the pulsars and non-pulsars correctly. In other words, it should have zero misidentified pulsars and zero misidentified non-pulsar candidates. This essence is conveyed in the accuracy, but, given the high-class imbalance in the dataset, the accuracy score is susceptible to be misleading. An `all-negative' classifier (which labels all the candidates as a non-pulsar irrespective of input data) would report an accuracy of $0.9$, if applied to  a dataset with $10$ pulsars (positive instances) and $90$ non-pulsars (negative instances), whereas recall, precision, $f_1$, G-Mean would all vanish. To alleviate this issue, we have employed SMOTE to artificially inflate the pulsar-set (positive instances) to make the dataset more balanced. We note that  the SMOTE dataset  led to no significant improvement in the scores compared to the non-SMOTE dataset, which is counterintuitive. 

\par One possible reason for the above result  could be that the tree-based algorithms are inherently more robust to class imbalance. Similar results have also  been observed by~\citet{Lyon14,Lyon}, wherein Gaussian Hellinger Very Fast Decision Tree (GH-VFDT) was shown to maximize the classification performance in heavily imbalanced data streams. In the case of an ANN (MLP), where the class imbalance is known to have a strong impact on its performance, we assert that our choice of learning rule and  amount of iterations for which we need to train, allowed the ANN (MLP) to overcome the high class-imbalance at the expense of large amount of training time and computational resources.

\par \bthis{Tree-based machine learning algorithms automatically yield `feature importances', which provide many insights about the data.  We have reported these feature importances along with mRMR based rankings on both the non-SMOTE and the SMOTE datasets (cf. Fig.~\ref{fig:feature_ranking}.)} Considering all the feature rankings, we found that the S/N of the folded profile ($f_{ID} = 0$) was the most descriptive feature. Note, that this is in agreement with similar conclusions from ~\citet{Lyon}.

\par In our set of machine learning methods, which we have successfully implemented, we note that recall, accuracy, and G-Mean varied  only after  the third significant digit. Therefore, for our dataset it was not possible to comparatively rank these algorithms on the basis of the above performance scores. 
So, we considered a  heuristic approach, based on qualitative aspects  to compare them. ANNs, which have been shown to work quite well on classification tasks such as this in earlier works provide no diagnostic information or interpretation of what is happening 'behind the scenes'. Interpretation of ANNs is a long standing problem in the Machine learning community, which renders ANNs as nothing more than black boxes (See~\citep{nnbb,nnbb2}, which try to provide some insight into the working of an ANN). Furthermore, they require huge amount of computational resources for training and testing, and offer a vast palette of hyper parameters, which makes it somewhat cumbersome to tune. While there are still some use-cases where, the ANNs and their variants , e.g. Convolutional Neural Networks (CNNs), are the state-of-the-art and unmatched by any other machine learning methods, the problem at hand is not one,  where the choice of an ANN would be ideal. Tree-based methods, on the other hand, seem very well suited as they are fast in training/testing, provide powerful insights about the data and are much simpler to train.
\par Adaboost, which is strictly not a tree-based ML method but a meta-classifier also does well on the problem at hand. It assigns and manipulates the weights of each datum in the training set so that even the `tricky` points are modeled and accounted for. One immediate concern this methodology raises is: how would it perform on noisy data? The scores reported may obfuscate the underlying requirement of clean, superior quality data for  the algorithm to train on, which unfortunately, is never at our disposal all the time.
\par The ideology of Gradient Boosting Classifier (GBC)  to sequentially train the classification trees on the errors of the predecessor tree in an additive fashion is almost identical to that of Adaboost. The main high level difference between the two algorithms is that, Adaboost assigns a high weight to the `tricky' datum, whereas GBC takes the help of another tree to learn that `tricky' datum. Quite interestingly, the idea of assigning weights was conceived before the idea of using the gradient, which is used to train the successor trees. Adaboost were first used in~\citeyear{Adaboost} and GBC in~\citeyear{gbc11}. Similar to the ANNs, the learning rate hyper parameter plays a decisive role in controlling the performance of GBC. One serious drawback of GBC is its sequential algorithm, which makes it powerless in the face of an enormous onslaught of data.
\par 
\rthis{We note that XGBoost}  supersedes Adaboost and  GBC, and provides a parallelization capability (including GPU support), has the ability of regularization, and is 
tailor-made for extreme value problems.  In the SKA era, which is akin to Big Data era for the radio astronomers, one would expect XGBoost to be the de-facto algorithm for most ML purposes.

We end, by  pointing out that our results and conclusions are based on a single dataset and feature set, used in ~\citet{Morello}, and we have not yet tested these algorithms on other datasets. 
So it is possible that our results are specific to the dataset been used. We do plan to implement these algorithms on other publicly available pulsar datasets (for eg.~\citealt{Lyon}) in future works. We also plan to apply other ML-based post-processing pipelines such as PICS~\citep{Zhu} on this same dataset.

We hope that  our results  based on publicly available post-processing data from the HTRU-S survey processed with the {\tt PEASOUP} pipeline provide impetus to the pulsar community to consider \rthis{implementing XGBoost as well as Adaboost and GBC} in their pulsar (and other radio transient search) pipelines to aid in the automated separation of RFI and other noise candidates. Conversely, we hope that this work spurs other pulsar groups to make their post-processing data public (similar to what is done for optical large-scale structure photometric and spectroscopic surveys), so that people outside the pulsar community can run and tune  their favorite machine learning algorithms on these datasets. We have also made available our ipython notebooks containing all the  codes to reproduce our results  at \url{https://github.com/shiningsurya/pulsarml}

\section*{Acknowledgments}
We would like to thank  Robert Lyon and Vincent Morello for invaluable feedback and comments on the manuscript, and to Vincent Morello and Ewan Barr for useful correspondence about the HTRU dataset and the SPINN algorithm, and for making their data public. We are also grateful to P.K. Srijith and  Ben Hoyle for many enlightening discussions about machine learning.



\bibliography{sample}






\label{lastpage}
\end{document}